\documentclass[conference]{IEEEtran}
\IEEEoverridecommandlockouts

\usepackage{cite}
\usepackage{amsmath,amssymb,amsfonts}
\usepackage{algorithmic}
\usepackage{graphicx}
\usepackage{textcomp}
\usepackage{xcolor}
\usepackage{subcaption}
\usepackage{multirow}
\usepackage{balance}
\usepackage{booktabs}
\def\BibTeX{{\rm B\kern-.05em{\sc i\kern-.025em b}\kern-.08em
    T\kern-.1667em\lower.7ex\hbox{E}\kern-.125emX}}

\begin{document}
\title{RT-VSS: In-House Ray Tracing for Vital Signs Sensing in FR3 Band%
\thanks{This research was conducted as part of the SMARTTEST project and has received funding from the European Union under Grant Agreement No. 101167834.  The work of Z. Cui is supported by the Research Foundation– Flanders (FWO),
Senior Postdoctoral Fellowship under Grant No. 12AFN26N.}
}
\author{\IEEEauthorblockN{Rudranil Chattopadhyay\IEEEauthorrefmark{1}\IEEEauthorrefmark{2}, Zhuangzhuang~Cui\IEEEauthorrefmark{1}\IEEEauthorrefmark{2}, Achiel Colpaert\IEEEauthorrefmark{1}\IEEEauthorrefmark{2}, and Sofie Pollin\IEEEauthorrefmark{1}\IEEEauthorrefmark{2}}
\IEEEauthorblockA{
\IEEEauthorrefmark{1}WaveCoRE, Department of Electrical Engineering (ESAT), KU Leuven, Belgium\\
\IEEEauthorrefmark{2}imec, Kapeldreef 75, 3001 Leuven, Belgium\\}
Email: \texttt{\{firstname.lastname\}@kuleuven.be}, \texttt{\{firstname.lastname\}@imec.be}}

\maketitle
\begin{abstract}
\label{abstract}
Vital Signs Sensing, enabled by the evolution of wireless communications, is inherently challenging as the chest's micro-Doppler signatures can be corrupted by multipath, body movement and noise. Additionally, the prospects of the Frequency Range (FR) 3 band, spanning 7-24 GHz, are compelling due to centimeter-scale wavelength improving phase sensitivity to chest motion, while its abundant bandwidth and rich multipath enhance temporal and spatial diversity. In this paper, we validate the feasibility of using cellular communications in the FR3 band to monitor a human's breathing and heartbeat in an indoor environment. Using Blender to model physiologically accurate vital-signs signals as ground truth and Sionna Ray Tracing simulations, we find that it is feasible to extract breathing and heartbeat waveforms at FR3 frequencies. Furthermore, we perform a comparative analysis of ISAC performance against representative sub-6 GHz (FR1) and mmWave (FR2) frequencies, indicating the utility of the study across the frequency bands.
\end{abstract}

\section{Introduction}
\label{introduction}
Non-intrusive vital signs monitoring has emerged as an important use-case of modern wireless systems, in the context of Integrated Sensing and Communication (ISAC)  \cite{ishabakaki, zhang_survey, wu_isac}. Wireless vital signs monitoring ensures comfort and privacy of human subjects, whilst allowing healthcare entities to tackle staff shortages. Several techniques and technologies have been studied, illustrating successful deployments, thus bridging healthcare and wireless communications \cite{li_nature}.

ISAC unifies sensing and communication, eliminating the need for dedicated infrastructure. Practical deployment strategies, especially for healthcare use-cases, require sufficient resources to support ISAC. The frequency range (FR) 3 band, spanning 7-24 GHz, emerges as a compelling solution and is expected to provide larger bandwidth (BW) than FR1  \cite{zach_fr3, bazzi_fr3}. Moreover, the sufficiently rich multipath environment allows for strategic beamforming capabilities, which become increasingly complicated at higher frequencies. Thus, the FR3 band becomes an integral component for ISAC.

Technologies like Radar \cite{liu_ncvs}, Camera \cite{Wijenayake2017}, WiFi \cite{wac}, ultra-wideband (UWB) \cite{ren_uwb}, etc., have enabled the detection and monitoring tracking of micro-Doppler variations in human subjects, eliminating the need to attach devices to the body \cite{kouhalvandi, radar_vss}. Such studies enhance modern healthcare by aiding diagnosis of disorders in breathing, heart, sleep, and more. 

While Radar offers impressive results in capturing the micro-Doppler effects, especially with mmWave Radars, it requires dedicated infrastructure, limiting scalability and investments for daily use \cite{kouhalvandi, radar_vss, mmrh, mercuri}. Cameras, on the other hand, raise concerns pertaining to privacy and requirement of ambient lighting and Line-of-Sight (LOS) conditions \cite{abuella}. ISAC-enabled architecture can reuse the existing communication infrastructure for sensing, even in non-LOS (nLOS) conditions owing to diversity. Additionally, OFDM waveforms used in 5G/6G and WiFi \cite{zhang_vss, hisac}, provide wideband channel data, readily available at receivers (RXs) to leverage for sensing.

A concern in utilizing channel data for various applications is the limited access to diverse datasets, especially for Machine Learning (ML)-based algorithms \cite{remcom, meneghello_dataset}. This can be attributed to the complications in planning measurement campaigns, the related finances, time and effort. To overcome these obstacles, simulation tools can be used, which not only boosts speed, modularity and scalability, but can also act as groundwork to ease setting up in-situ measurements.

Motivated by the existing challenges, this paper proposes a simulation-based framework for vital signs estimation in the FR3 band using Blender and Sionna's Ray Tracing (RT) tool. The contributions of this paper are as follows:
\begin{enumerate}
    \item Constructing a multipath-rich indoor scene with physiologically accurate human with breathing and heartbeat
    \item Utilizing signal processing to extract vital signs from a RT-based simulation channel data
    \item Comparing FR3 sensing and communication performance against representative FR1 and FR2 frequencies
\end{enumerate}

The paper is organized as follows: Sec. \ref{vss} presents the procedure to build the RT scene in Blender, modeling of vital signs, configuration of the simulation setup and data collection. In Sec. \ref{sp}, the signal processing steps to extract vital signs are explained. In Sec. \ref{sensing}, the sensing performance across FR1, FR2 and FR3 bands is presented. In Sec. \ref{comms}, the communication performance and impact of chest motion on communication links is highlighted. Finally, Sec. \ref{conclusion} concludes the paper with some insights into prospective work.

\section{Vital Signs Simulation Framework}
\label{vss}
The simulation framework proposed in this paper has two phases.
\subsection{Scene Construction}
\begin{figure}[htbp]
\centering
\begin{subfigure}[b]{0.49\linewidth}
    \centering
    \includegraphics[width=\linewidth, trim =  4cm 0cm 1cm 0cm, clip]{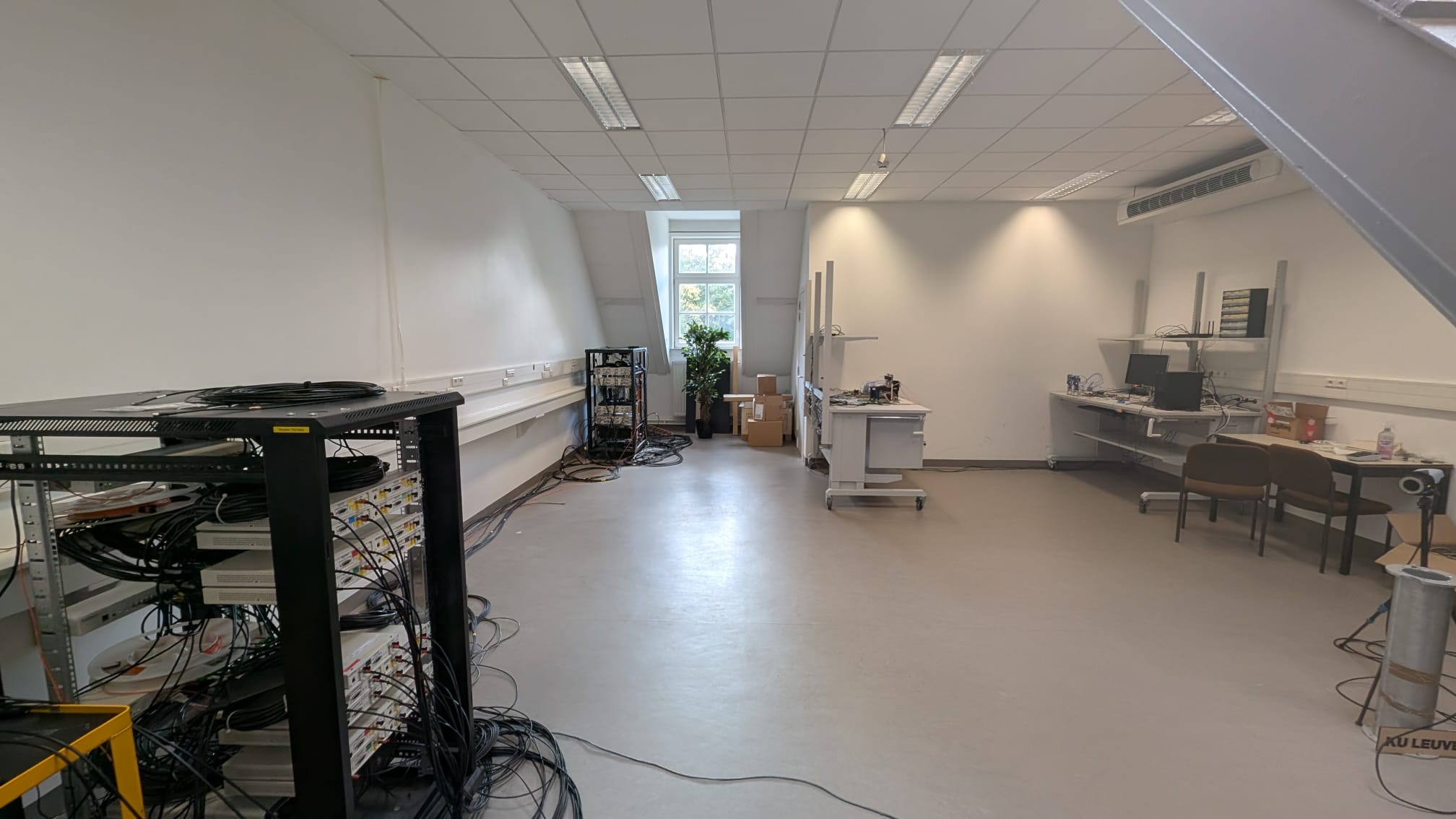}
    \caption{Real-world Layout.}
    \label{OG_MIMO}
\end{subfigure}
\hfill
\begin{subfigure}[b]{0.49\linewidth}
    \centering
    \includegraphics[width=\linewidth, trim =  1cm 5cm 1cm 1cm, clip]{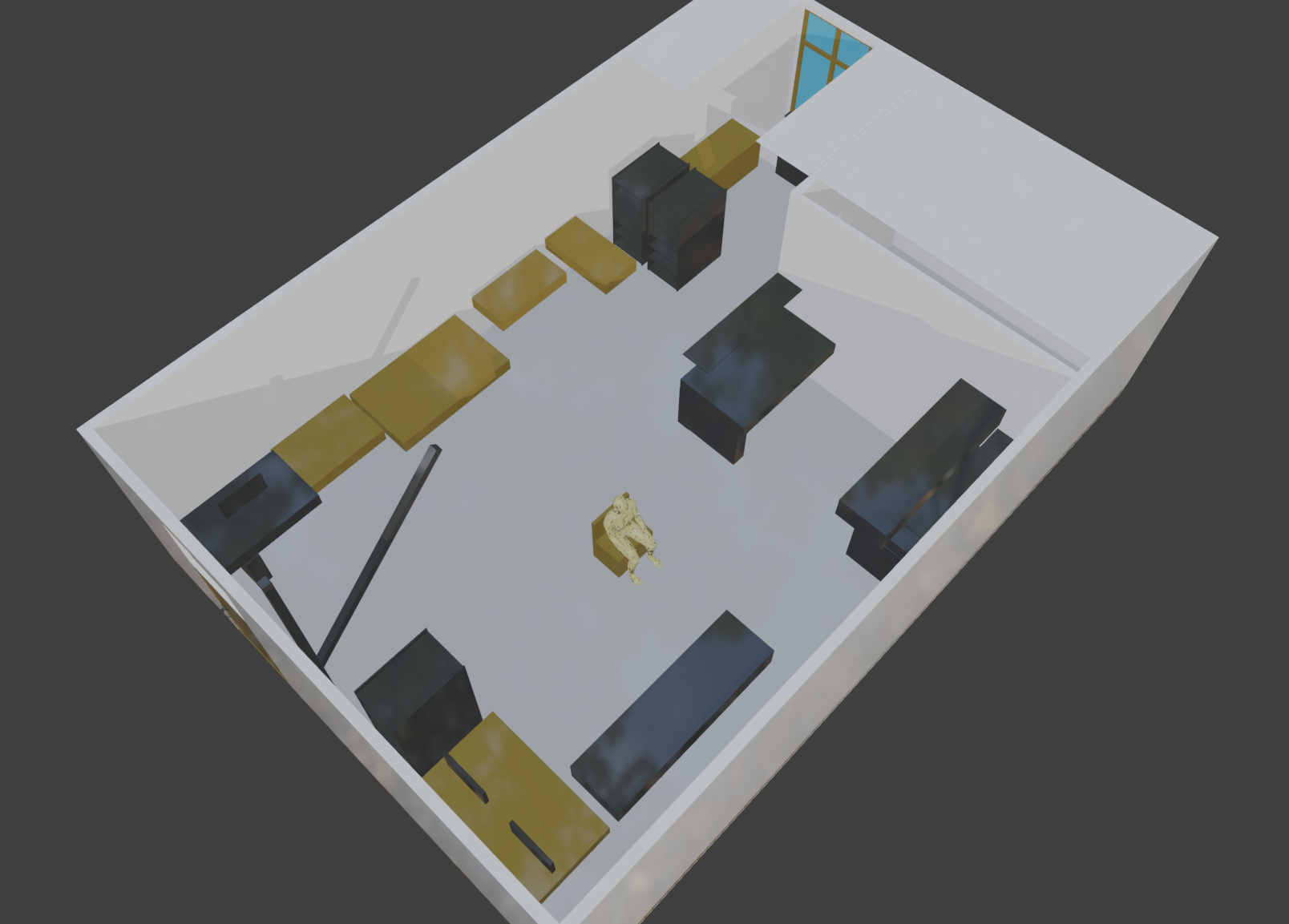}
    \caption{Blender Model.}
    \label{MIMO_Model}
\end{subfigure}
\caption{(a) Real-world and (b) replicated Blender model of MIMO Lab at WaveCoRe, ESAT, KU Leuven.}
\end{figure}

The simulation scene is an indoor laboratory, which replicates the real-world setup of the "MIMO Lab" in the ESAT WaveCoRe Group at KU Leuven, shown in Figs.\ref{OG_MIMO} and \ref{MIMO_Model}. The digital model of the Lab is built using Blender 4.2LTS \cite{blender}, with dimensions $7.6 \times 10.5 \times 3.1$ m\textsuperscript{3}. The Lab consists of different furniture and devices, built with different materials \cite{sionnart_report, ITURP2040_3, site_fr3}, providing a rich multipath environment.

The modeling and animation of human model in Blender is adapted from \cite{remcom}, validating the methodology of the Blender-RT framework. While \cite{remcom} measures radar backscatter in a semi-anechoic chamber using the proprietary WaveFarer tool, the agreement of simulated and measured phase variations proves feasibility of physiologically-grounded simulation.

In this work, the human mesh is generated using makeHuman-community tool \cite{makehuman}. The model is not constrained by age, gender, ethnicity or other physical features, but represents an average person. The human is in a sitting posture and is extracted as a \textit{.fbx} element. The human model is imported in the Blender scene and placed at [4.0, 4.0]. The chest is at a height of 0.87 m from the ground.

The human mesh was decimated to save computation effort, producing a mesh characterized by 8,026 triangular primitives and 4,122 vertices, representing the skin with frequency-dependent-EM parameters (in Tab. \ref{em}) \cite{Gabriel1996}. The scattering behavior of EM waves on the human body is characterized using the Beckmann-Kirchhoff surface scattering theory \cite{beckmann} as an engineering approximation. This explains when a plane wave is incident on a surface, it is reflected as specular and diffuse reflections. Assuming the RMS height ($\sigma_h$) of human skin is $\approx$ 0.21 mm \cite{diffuse}, the scattering coefficient \textit{S} at frequency $f_c$ is calculated using Eq. \ref{scattering_coefficient}:
\begin{equation}
S = \sqrt{1 - \exp\left(-\left(\frac{4\pi\cdot\sigma_{h}\cdot f_{c}}{c}\right)^2\right)}
\label{scattering_coefficient}
\end{equation}
\begin{table}[htbp]
\caption{EM Properties of human skin as a function of frequency.}
\begin{center}
\begin{tabular}{ccccc}
\toprule
\textbf{Band} &
\textbf{Frequency} &
\textbf{\textit{Relative}} &
\textbf{\textit{Conductivity}} &
\textbf{\textit{Scattering}} \\
&
\textbf{[GHz]} &
\textbf{\textit{Permittivity}} &
\textbf{\textit{[S/m]}} &
\textbf{\textit{Coefficient}} \\
\midrule

FR1 & 6  & 38.378 & 4.5421 & 0.053 \\
\midrule

\multirow{6}{*}{FR3}
 & 7  & 37.146 & 5.5823 & 0.062 \\
 & 10 & 33.528 & 8.9510 & 0.089 \\
 & 12 & 31.252 & 11.260 & 0.107 \\
 & 18 & 25.384 & 17.772 & 0.160 \\
 & 20 & 23.765 & 19.708 & 0.178 \\
 & 24 & 20.980 & 23.192 & 0.213 \\
\midrule

FR2 & 28 & 18.714 & 26.187 & 0.248 \\
\bottomrule

\end{tabular}
\label{em}
\end{center}
\end{table}

\begin{figure}[htbp]
\centering
\begin{subfigure}[b]{0.48\linewidth}
    \centering
    \includegraphics[width=\linewidth]{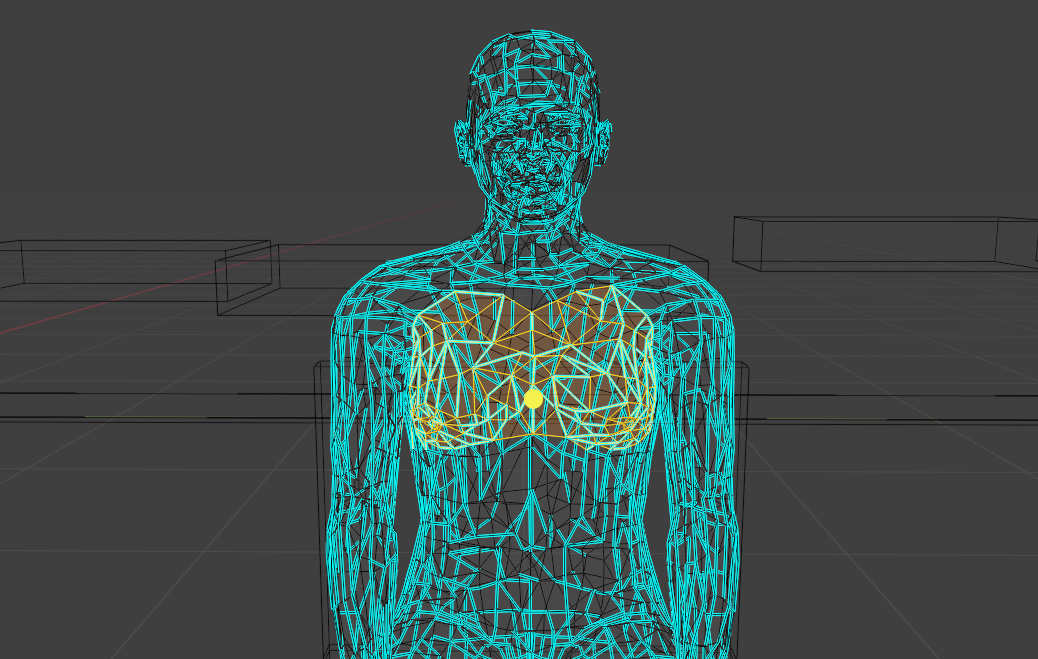}
    \caption{Vertices for Breathing.}
    \label{breathing_blender}
\end{subfigure}
\hfill
\begin{subfigure}[b]{0.48\linewidth}
    \centering
    \includegraphics[width=\linewidth]{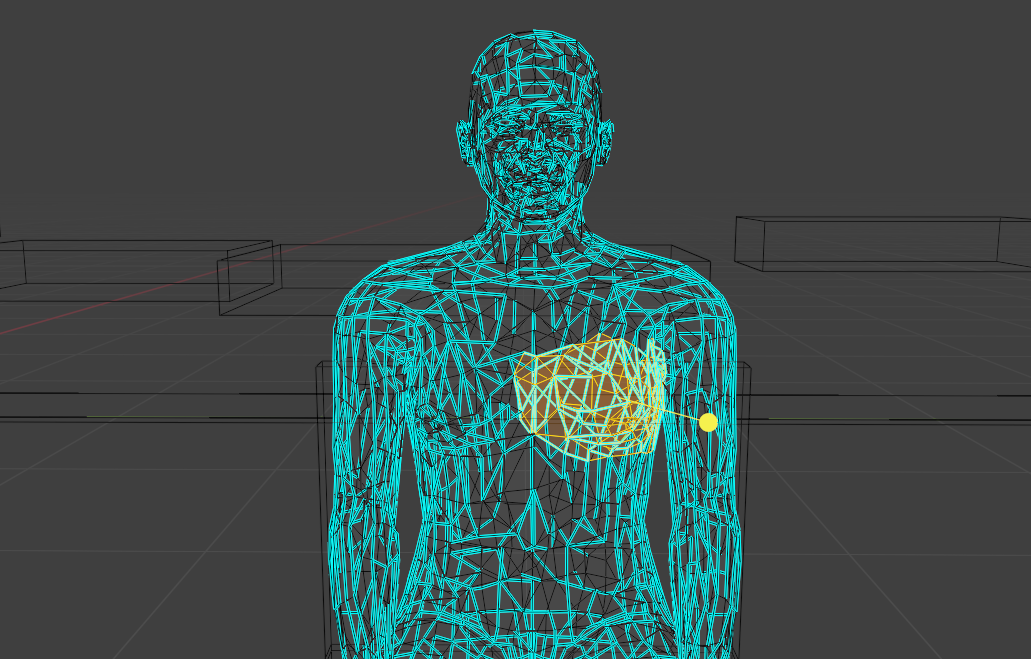}
    \caption{Vertices for Heartbeat.}
    \label{heartbeat_blender}
\end{subfigure}
\caption{Selected vertices to emulate (a) breathing and (b) heartbeat in the human model in Blender.}
\end{figure}

\begin{figure}[htbp]
\centerline{\includegraphics[scale = 0.5]{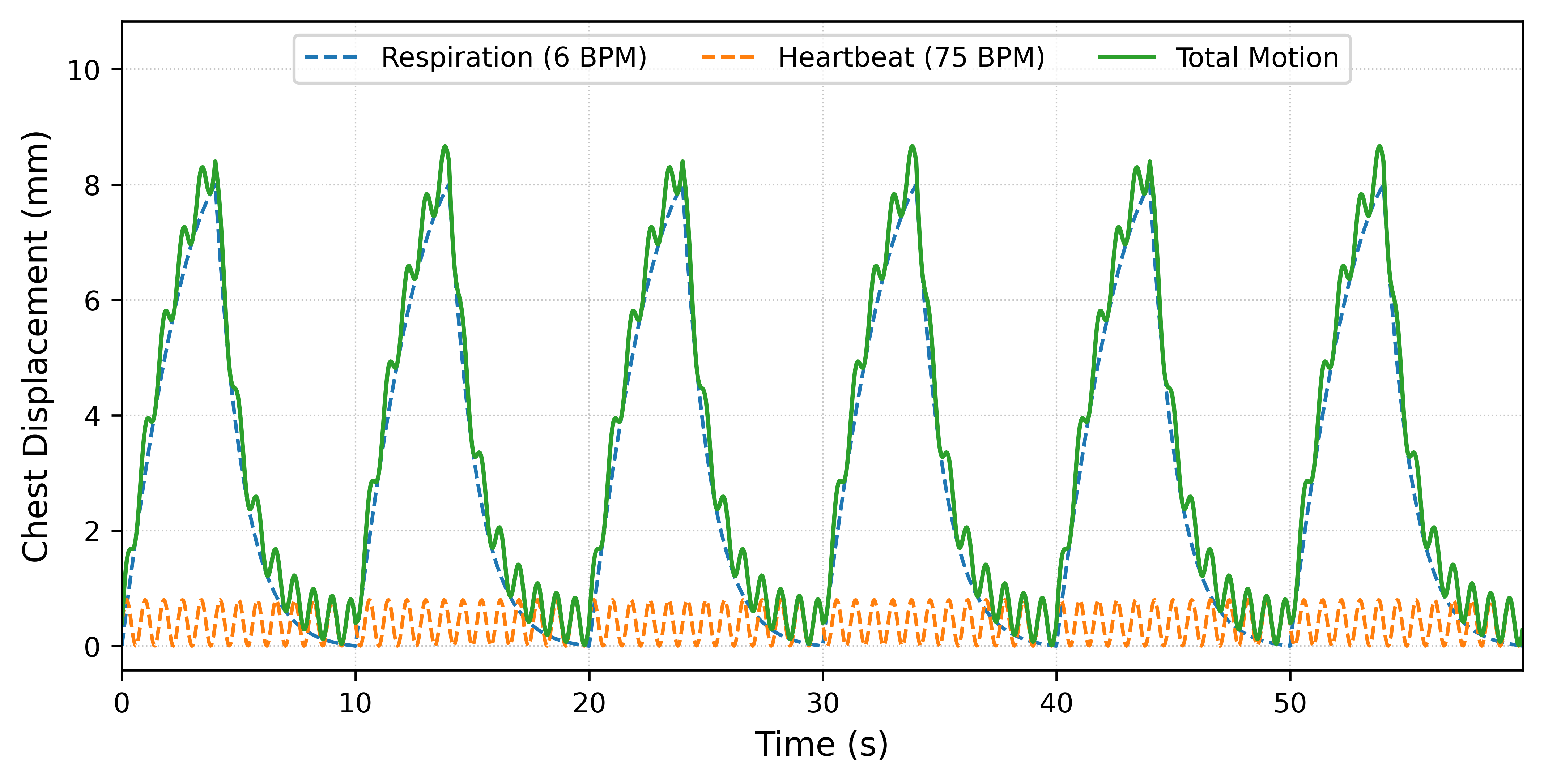}}
\caption{Ground truth chest motion for breathing and heartbeat.}
\label{gt}
\end{figure}

These vertices are grouped and assigned functions using shape keys feature in Blender to emulate respiration and heartbeat \cite{remcom} (Figs. \ref{breathing_blender} and \ref{heartbeat_blender}). The vertices corresponding to the chest were selected and assigned parabolic inhalation and exponential exhalation with Eqs. \ref{inhalation_gt} and \ref{exhalation_gt} respectively \cite{albanese}:
\begin{equation}
d_{\text{inhalation}}(t)
=
\frac{-d_{\text{b,max}}}{T_{i} \cdot T_{e}} t^2
+
\frac{d_{\text{b,max}}}
{T_{i} \cdot T_{e} \cdot f_{\text{breath}}}
t
\label{inhalation_gt}
\end{equation}
\begin{equation}
d_{\text{exhalation}}(t) = \frac{d_{\text{b,max}}}{1 - e^{\frac{-T_{e}}{\tau}}} \left( e^{\frac{-(t - T_{i})}{\tau}} - e^{\frac{-T_{e}}{\tau}} \right)
\label{exhalation_gt}
\end{equation}
where, $d_{\text{inhalation}}(t)$ and $d_{\text{exhalation}}(t)$ are chest motion at time $t$ during respiration, $d_{\text{b,max}}$ is maximum chest displacement, $f_{\text{breath}}$ is breath rate (BR), $T_{i}$ and $T_{e}$ are inhalation and exhalation times in one breath cycle and $\tau$ is exhalation decay constant. Similarly, vertices for the heart region were selected and assigned heartbeat with Eq. \ref{heart_gt}:
\begin{equation}
d_{\text{heart}}(t) = d_{\text{h,max}} \cdot\sin\left( \frac{2\pi \cdot t\cdot f_\text{heart}}{60} \right) + d_{\text{h,max}}
\label{heart_gt}
\end{equation}
where, $d_{\text{heart}}(t)$ is chest motion at time $t$ due to heartbeat, $d_{\text{h,max}}$ is maximum chest displacement due to heartbeat, and $f_{\text{heart}}$ is heart rate (HR).
The total chest displacement ($d_{\text{total}}$) is given by Eq. \ref{total_gt}:
\begin{equation}
d_{\text{total}}(t) = d_{\text{breath}}(t) + d_{\text{heart}}(t)
\label{total_gt}
\end{equation}
As a systemic choice to obtain spectrally clean signals, the human is assumed to be in a calm state. The animated human has a BR of 6 beats per minute (BPM) (0.1 Hz) and HR of 75 BPM (1.25 Hz), which serves as the ground-truth for the simulations in this paper, visualized in Fig. \ref{gt}.

Having modeled and animated the scene, the next step is to export the 60-second-long-animation with continuous breathing and heartbeat to Sionna RT \cite{sionnart}. At 24 frames per second (FPS), 1,440 frames are exported as individual scenes, using Mitsuba XML extension in Blender \cite{mitsuba}. It is imperative to mention that each frame must have its own mesh files to capture the animation accurately.
\subsection{RT Scene Setup}
The simulation has one Transmitter (TX) with $P_{\text{TX}} = 0$ \text{dBm} and six RXs, shown in Tab. \ref{devices}. Out of six RXs, there are two dedicated sensing nodes co-located with TX, placed 10 cm apart from TX on each side. The carrier frequency is selected among: 6, 7, 10, 12, 18, 20, 24 and 28 GHz. A BW of 100 MHz is used for FR1 and 400 MHz is used for FR2 and FR3 frequencies \cite{zach_fr3}.
\begin{table}[htbp]
\caption{Radio Devices usage and location in the RT scene.}
\begin{center}
\begin{tabular}{ccc}
\toprule
\textbf{Radio Device}&\textbf{Purpose}&{\textbf{{Location [x, y, z]}}} \\
\midrule
TX & Transmitting Node & [5.0, 4.0, 0.85] \\
\midrule
RX 1 & Sensing Node & [5.0, 3.9, 0.85] \\
RX 2 & Sensing Node & [5.0, 4.1, 0.85] \\
\midrule
RX 3 & Communication Node & [2.2, 2.4, 0.85] \\
RX 4 & Communication Node & [3.1, 5.5, 0.85] \\
RX 5 & Communication Node & [5.8, 5.2, 0.85] \\
RX 6 & Communication Node & [6.0, 3.6, 0.85] \\
\bottomrule
\end{tabular}
\label{devices}
\end{center}
\end{table}

\begin{figure}[htbp]
\centerline{\includegraphics[scale = 0.3, trim =  0cm 2cm 0cm 1cm, clip]{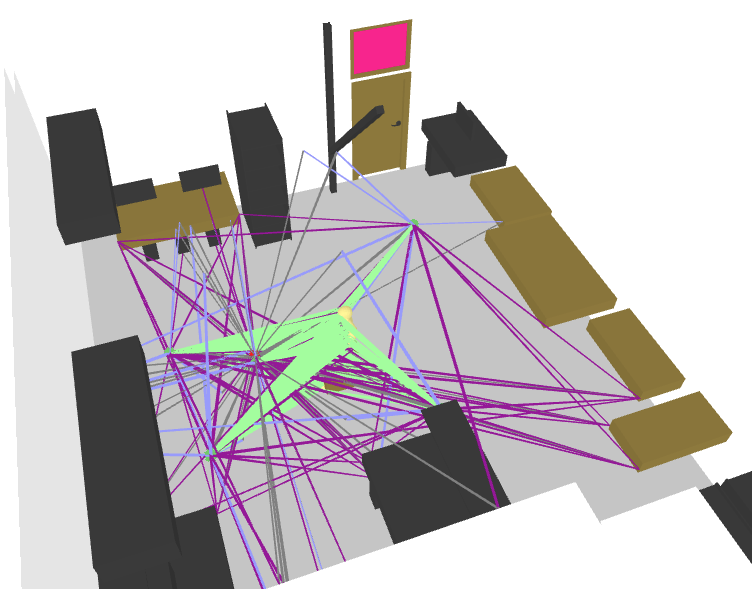}}
\caption{Ray Tracing in MIMO Lab in SionnaRT.}
\label{sionna_env}
\end{figure}

The rays can undergo a maximum of three surface interactions (\textit{max\_depth}) in the scene. Fig. \ref{sionna_env} shows rays for different propagation effects (\textit{max\_depth} = 1 for easier distinction): specular reflection (blue), diffuse reflection (green), refraction (orange) and diffraction (purple).

In Sionna, RT is performed and data is collected for each of the 1,440 frames. Each frame's XML file and meshes are loaded as scene, the TX and RXs are added and the Path Solver is initialized, which computes and traces the probable propagation paths in the scene. The frequency, BW, and EM Parameters (relative permittivity, conductivity and scattering coefficient) of human skin are assigned and RT is performed. The Paths object, instantiated by Path Solver, generates the channel data and is stored as \textit{.npz} structures.

\section{Signal Processing for Vital Signs Estimation}
\label{sp}
The vital signs information are extracted from the channel utilizing signal processing techniques. The signal processing algorithm is applied to the BW-limited channel. These steps are similar for both breathing and heartbeat extraction.
\subsection{Range Bin Selection}
The delay bin corresponding to the chest reflection is identified first. The bins in the range/delay profile are scored based on the product of mean amplitude and power in vital signs-bands. The bin scoring function is defined in Eq. \ref{range_bin}:
\begin{equation}
\text{score}_k = P_k^{\text{VS}} \cdot \bar{A}_k \cdot \mathbf{1}\!\left[\bar{A}_k > 0.1 \cdot \max_j \bar{A}_j\right]
\label{range_bin}
\end{equation}
where $P_k^{\text{VS}}$ is the vital-signs-band spectral power at bin $k$. The indicator term $\mathbf{1}[\cdot]$ suppresses bins whose mean amplitude $\bar{A}_k$ falls below 10\% of the strongest bin, preventing selection of noisy components. The highest-scoring-bin is selected and its variations are tracked across all frames. The optimal bin is:
\begin{equation}
k^* = \underset{k}{\arg\max}\ \!\left[\text{score}_k\right]
\end{equation}


\subsection{Extraction of Phase}
In this work, the variations in phase across subsequent frames throughout the measurement period is utilized. The phase variations capture the micro-Doppler due to the displacement caused by the breathing and heartbeat of the human body. The complex time series $S[k^*, :] \in \mathbb{C}^{N_{\text{frames}}}$ obtained from the previous step is used to extract phase using Eq. \ref{phase}:.
\begin{equation}
    \phi = \frac{4\pi \cdot f_c \cdot \Delta r}{c} \quad [\text{rad/m}]
    \label{phase}
\end{equation}
where, $c$ is the speed of light and $\Delta$r is the chest displacement.

The raw phase $\phi_{\text{wrapped}}[n] = \angle\, S[k^*, n]$ is bounded by $(-\pi, +\pi]$. 
The phase excursion crosses the $\pm\pi$ boundary multiple times, for example, at 20 GHz a displacement of 8 mm causes a phase swing of $\approx$ 6.7 rad. Phase "unwrapping" recovers the true continuous phase trajectory by detecting consecutive differences exceeding $\pi$. This unwrapped signal is a combination of phase variations caused due to the environment, noise, breathing and heartbeat.
\begin{equation}
    \Phi_{\text{unwrapped}}(t) = \Phi_{\text{static}} + \Phi_{\text{breathing}}(t) 
+ \Phi_{\text{heartbeat}}(t) + \Phi_{\text{noise}}(t)
\end{equation}
where, $\Phi_{\text{unwrapped}}(t)$ is the total phase signal and $\Phi_{\text{static}}$, $\Phi_{\text{breathing}}(t)$, $\Phi_{\text{heartbeat}}(t)$ and $\Phi_{\text{noise}}(t)$ are phase variations due to propagation, breathing, heartbeat and noise, respectively.

The DC component, corresponding to the phase variation due to EM wave propagation and noise across the frames, is removed. Finally, the AC signal ($ \Phi_{\text{AC}}(t)$) is obtained which corresponds to the variations caused by the vital signs.
\begin{equation}
    \Phi_{\text{AC}}(t) \approx \Phi_{\text{breathing}}(t) + \Phi_{\text{heartbeat}}(t)
\end{equation}
\subsection{Filtering and FFT}
A 3\textsuperscript{rd}-order and a 4\textsuperscript{th}-order Butterworth filter is applied for BR and HR extraction respectively. The Butterworth filter is chosen because it has a maximally flat passband - no ripple within the breathing band - which is important to extract the asymmetric waveform. For breathing, the band 0.05 - 0.5 Hz and for heartbeat, the band 1.0 - 2.0 Hz are considered, ranging between 3 - 30 BPM and 60-120 BPM respectively.
\section{Sensing Performance}
\label{sensing}
\begin{figure}[htbp]
\centerline{\includegraphics[scale = 0.18]{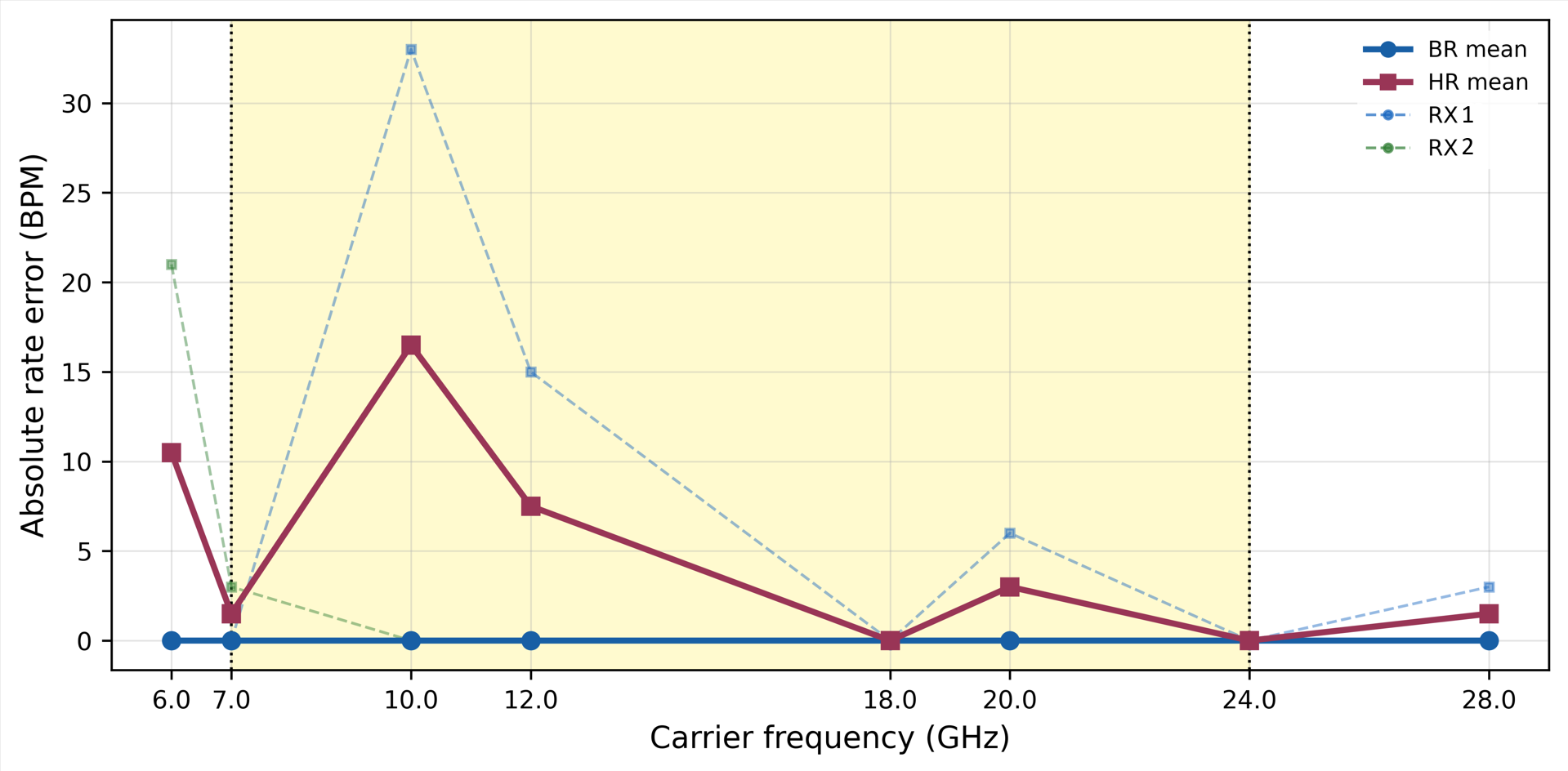}}
\caption{Impact of geometry and location of Sensing Nodes (RXs 1 and 2) on estimation of BR and HR Error Rates.}
\label{errors}
\end{figure}

In this work, two RXs (RX1 and RX2) only perform sensing. Fig. \ref{errors} shows the BR and HR errors in FR3 band and representative frequencies of FR1 and FR2 bands. The red and dark blue solid curves correspond to the HR and BR errors. The light blue and green dashed curves correspond to RX1 and RX2. BRs are shown with circles and HRs with squares. It can be observed that both RXs accurately estimate the BRs at 6 BPM. RX2 fails to accurately detect HRs below $f_c = $12 GHz, due to small amplitude of heartbeat, otherwise detecting at 75 BPM. RX1, however, fails to detect the HR accurately even at some high frequencies, with errors up to 32 BPM, unacceptable in clinical standards \cite{liebetruth}. The variation in performance of the RXs can be attributed to differences in respective geometric and propagation perspectives. This shows that some frequencies fail at certain positions. Hence, carefully planned deployment is essential. It can be quite interesting to study if using different frequencies (multi-band) for breathing and heartbeat detection can aid in improving sensing performance, due to larger available BWs offering better resolution for each case. 

\begin{figure}[htbp]
\centerline{\includegraphics[scale = 0.5]{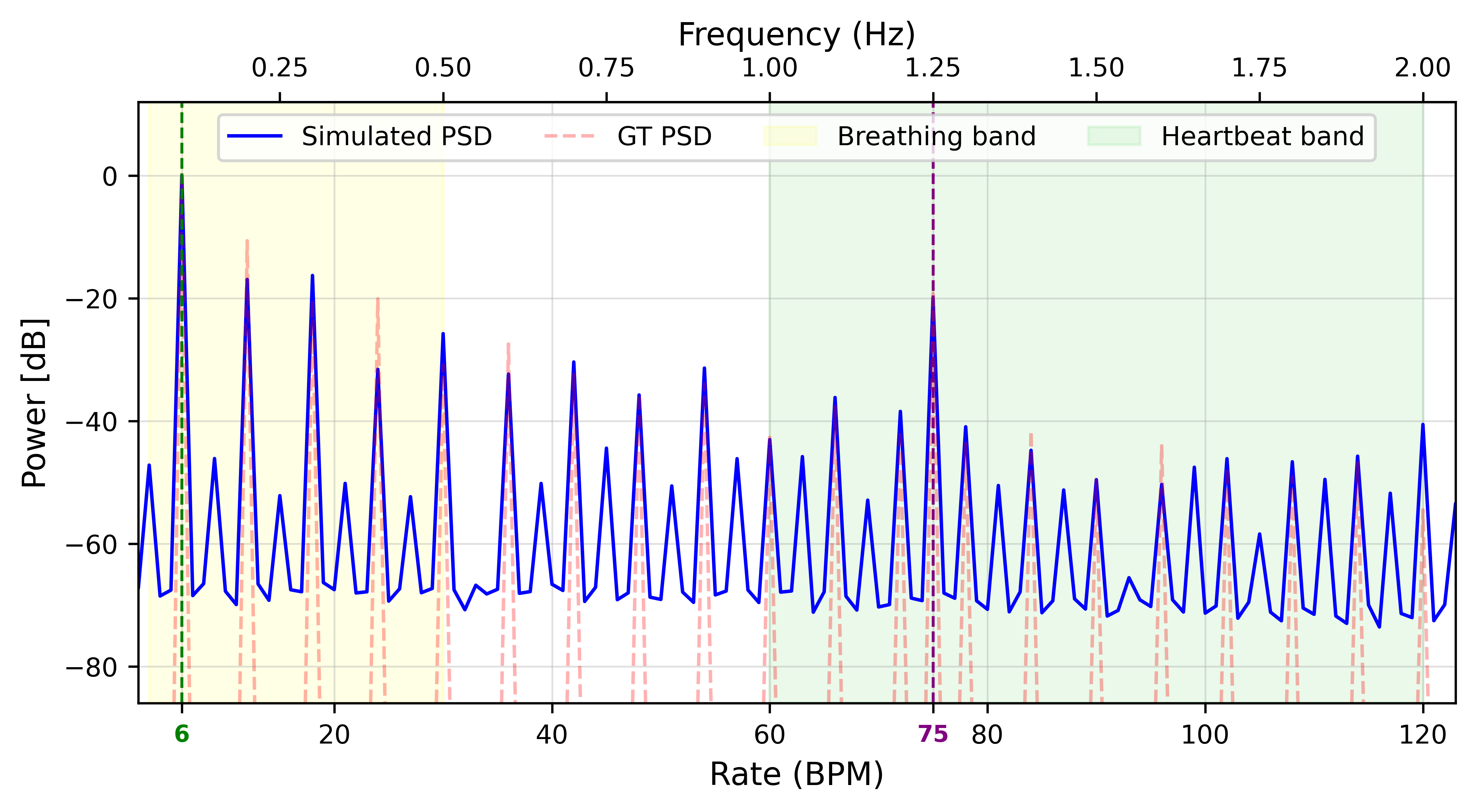}}
\caption{PSD at 20 GHz showing 0.1 Hz BR and 1.25 Hz HR.}
\label{FFT}
\end{figure}

Fig. \ref{FFT} shows the power spectral density (PSD) vs rate (bottom x-axis) and frequency (top x-axis) for RX2 at 20 GHz. The yellow and green regions highlight the breathing and heartbeat bands. It can be seen that the highest peaks are accurately detected at 6 and 75 BPM - the true BR and HR.

\begin{figure}[htbp]
\centerline{\includegraphics[scale = 0.5]{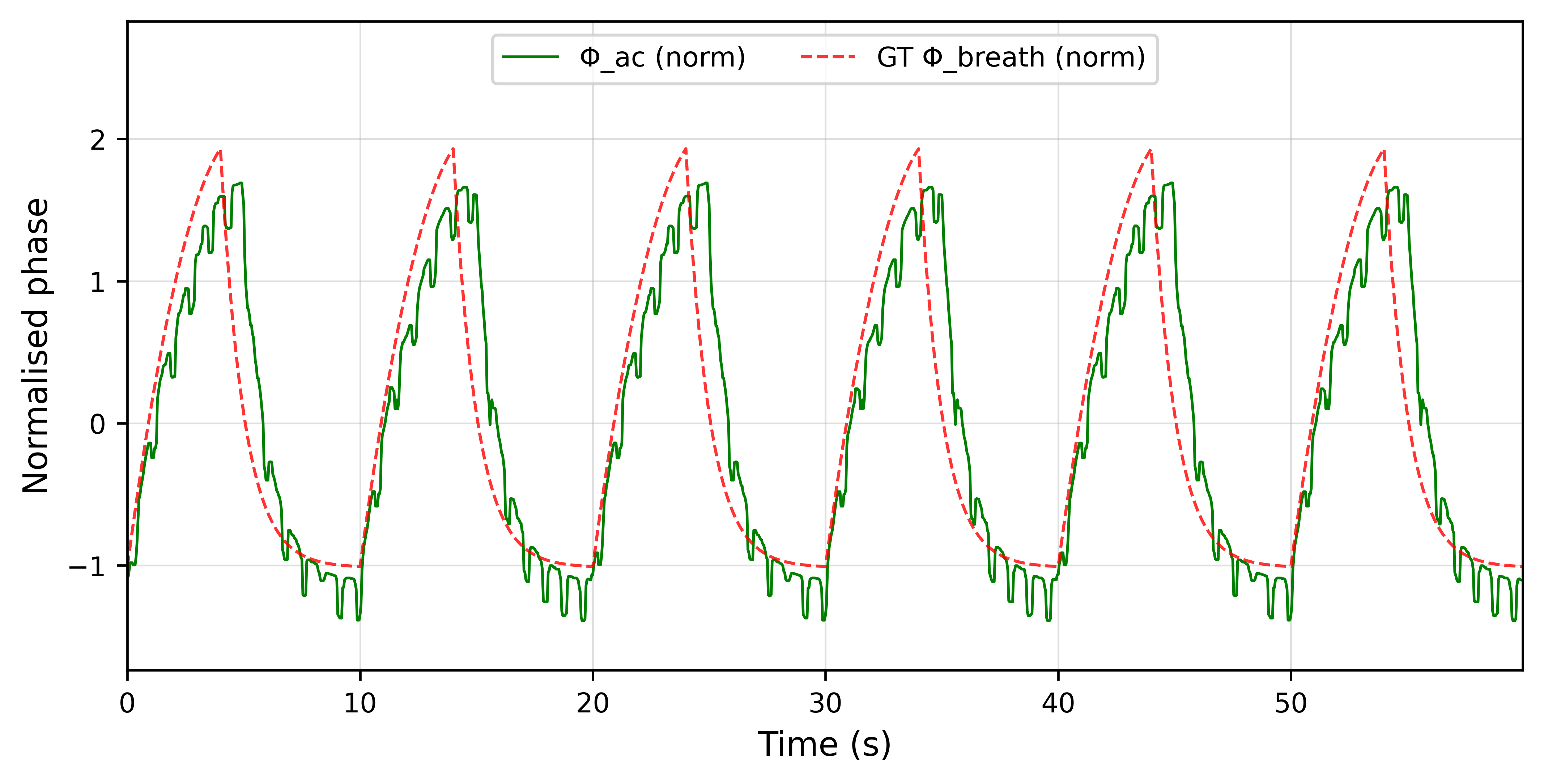}}
\caption{Breathing waveform extracted from channel at 20 GHz.}
\label{breathing_signal}
\end{figure}

At 20 GHz for RX2, the breathing and heartbeat waveforms are extracted from phase variations. Fig. \ref{breathing_signal} shows the extracted breathing signal. The extracted breathing signal closely follows the periodic ground-truth signal. The perturbations in the extracted signal can be attributed to the variations caused by heartbeat superposed on the breathing signal.

\begin{figure}[htbp]
\centerline{\includegraphics[scale = 0.5]{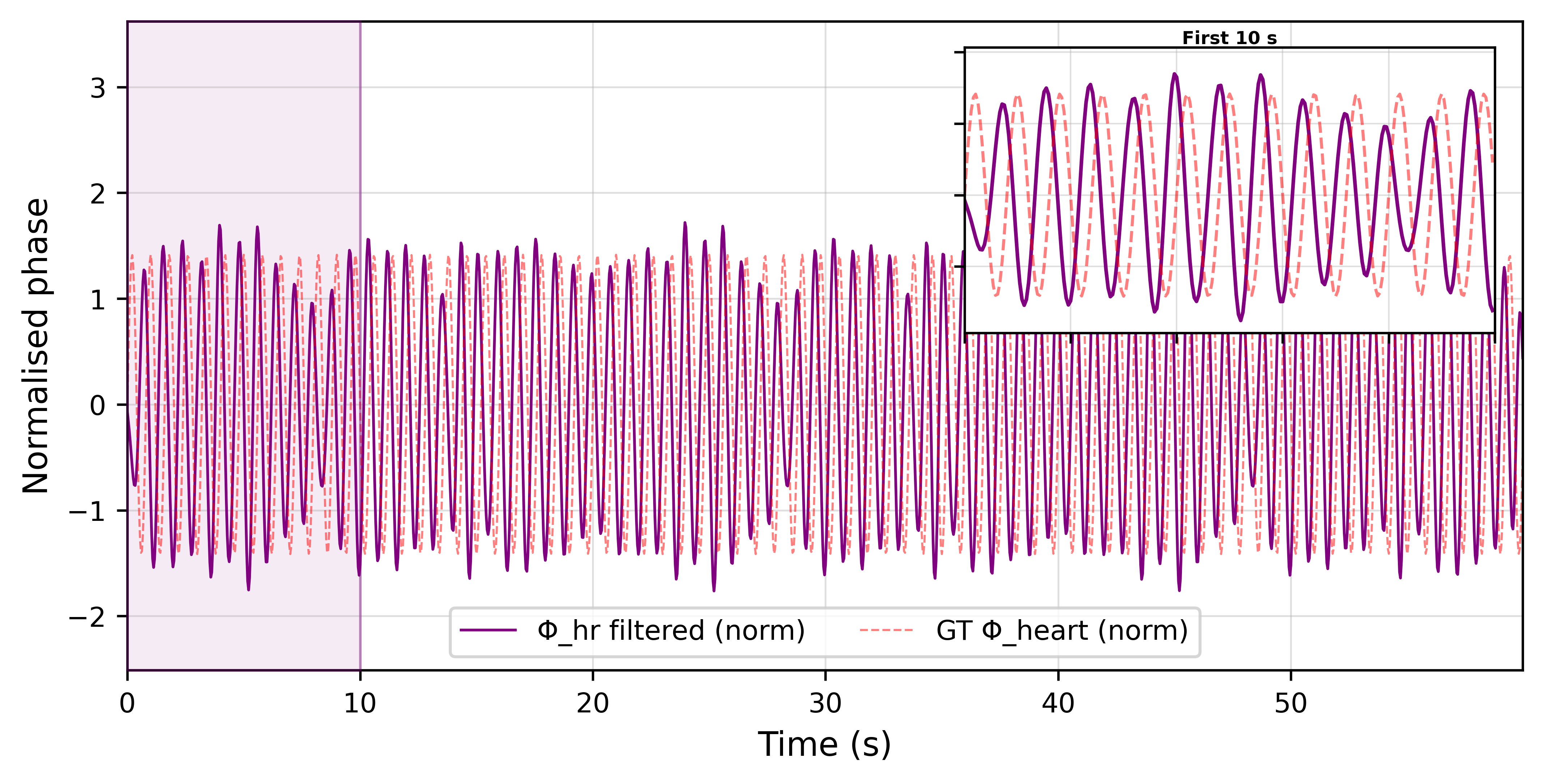}}
\caption{Heartbeat waveform extracted from channel at 20 GHz.}
\label{heartbeat_signal}
\end{figure}

Fig. \ref{heartbeat_signal} shows the ground-truth and extracted heartbeat signal, alongwith a snippet of the first 10 seconds of measurement. The simulation waveform is quite similar to the ground-truth waveform. However, there exists misalignment, which can be attributed to its low magnitude and coupling with breathing harmonics, which causes the distortions.
\section{Communication Performance}
\label{comms}

In this section, RXs used for communication are discussed. With one TX in the scene, resources are shared among sensing and communication nodes. The impact of chest motion on the communication links is analyzed using Path Loss (PL) and Spectral Efficiency (SE). To perform comparison, RT is performed and channel data is acquired for three different scenes, without varying other parameters:
\begin{enumerate}
    \item No Human: Only the lab environment with devices and furniture as scatterers, and no human
    \item Static: The human is present without any chest motion
    \item Vitals: Human chest motion averaged over 1,440 frames
\end{enumerate}
Average PL across all the communication links at the communication nodes in the scene is computed with Eq. \ref{PL_Eq}:
\begin{equation}
    PL_{\text{overall}} = - 10\log_{10}\left(\mathbb{E}\left[|H|^2\right]\right)
    \label{PL_Eq}
\end{equation}
where, expectation $\mathbb{E}[\cdot]$ is taken over all subcarriers. Average SE across all the communication nodes is computed using the Shannon formula using Eq. \ref{SE_Eq}:
\begin{equation}
    SE_{\text{overall}} = \mathbb{E}\left[\log_2\left(1 + \text{SNR}_k\right)\right]
    \label{SE_Eq}
\end{equation}
where, $\text{SNR}_k$ is per-subcarrier SNR with thermal noise power in per-subcarrier BW.

\begin{figure}[htbp]
    \centering
    \includegraphics[width=\columnwidth]{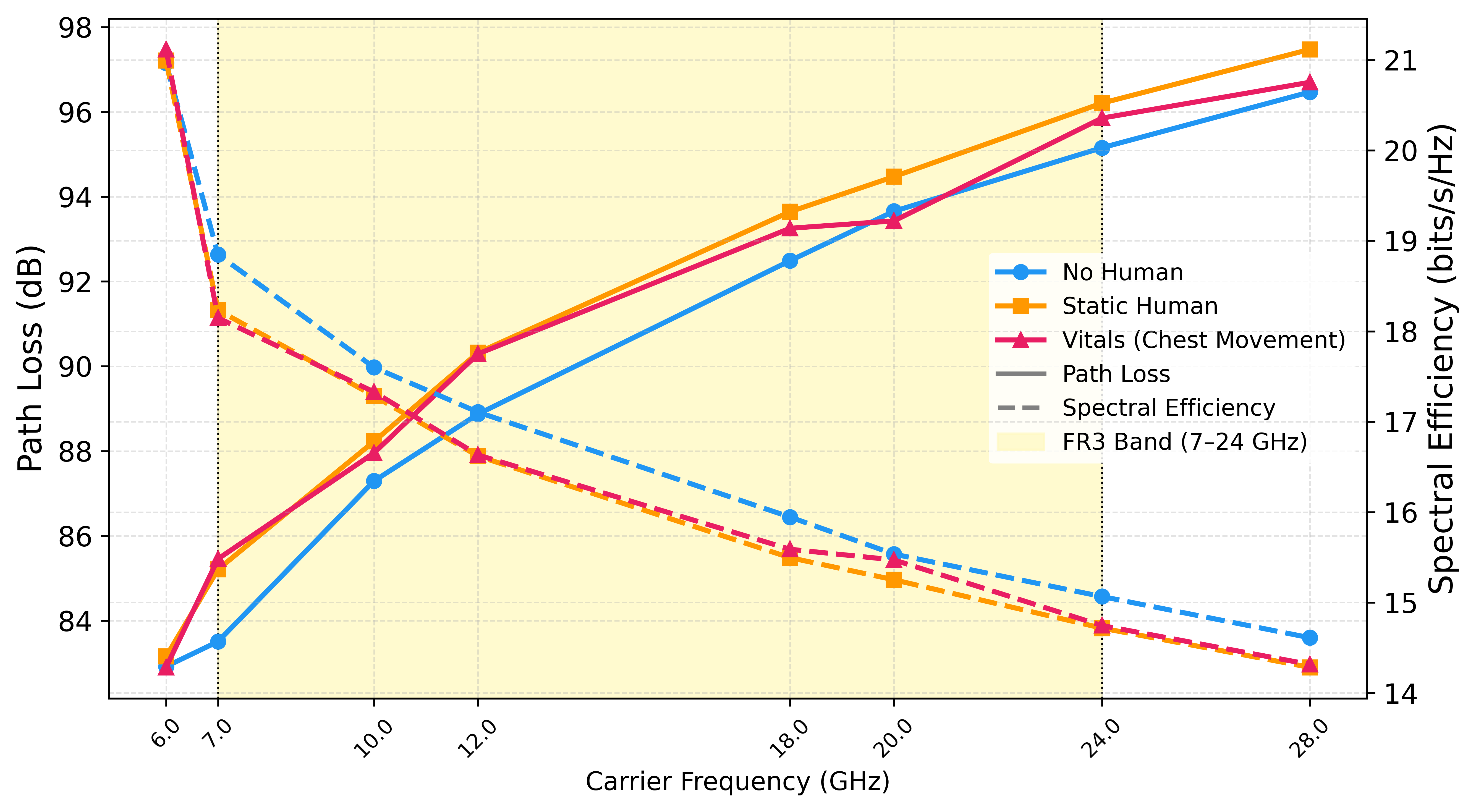}
    \caption{Variation of PL and SE with frequency, across communication nodes (RXs 3 - 6) at 20 GHz.}
    \label{PL_SE}
\end{figure}

Fig. \ref{PL_SE} shows the overall performance of communication links across all RXs. It can be observed that addition of human to the scene increases PL by $\approx$ 1-1.5 dB across the frequencies, owing to absorption by human. However, at higher frequencies, the "Vitals" curve shows lower PL than "Static" curve, indicating that in such cases, chest motion creates paths interfering constructively, leading to better performance than "Static". Compared to the "No Human" scene, SE for "Vitals" scene reduces by a maximum of $\approx$ 0.5 bits/s/Hz. The static scene shows worse performance, especially in FR3 and FR2 bands. These indicate that while presence of humans affects the communication links, at higher frequencies, on average, chest motion can benefit communication performance.

\begin{figure}[!t]
    \centering

    \begin{subfigure}[t]{0.49\columnwidth}
        \centering
        \includegraphics[width=\linewidth]{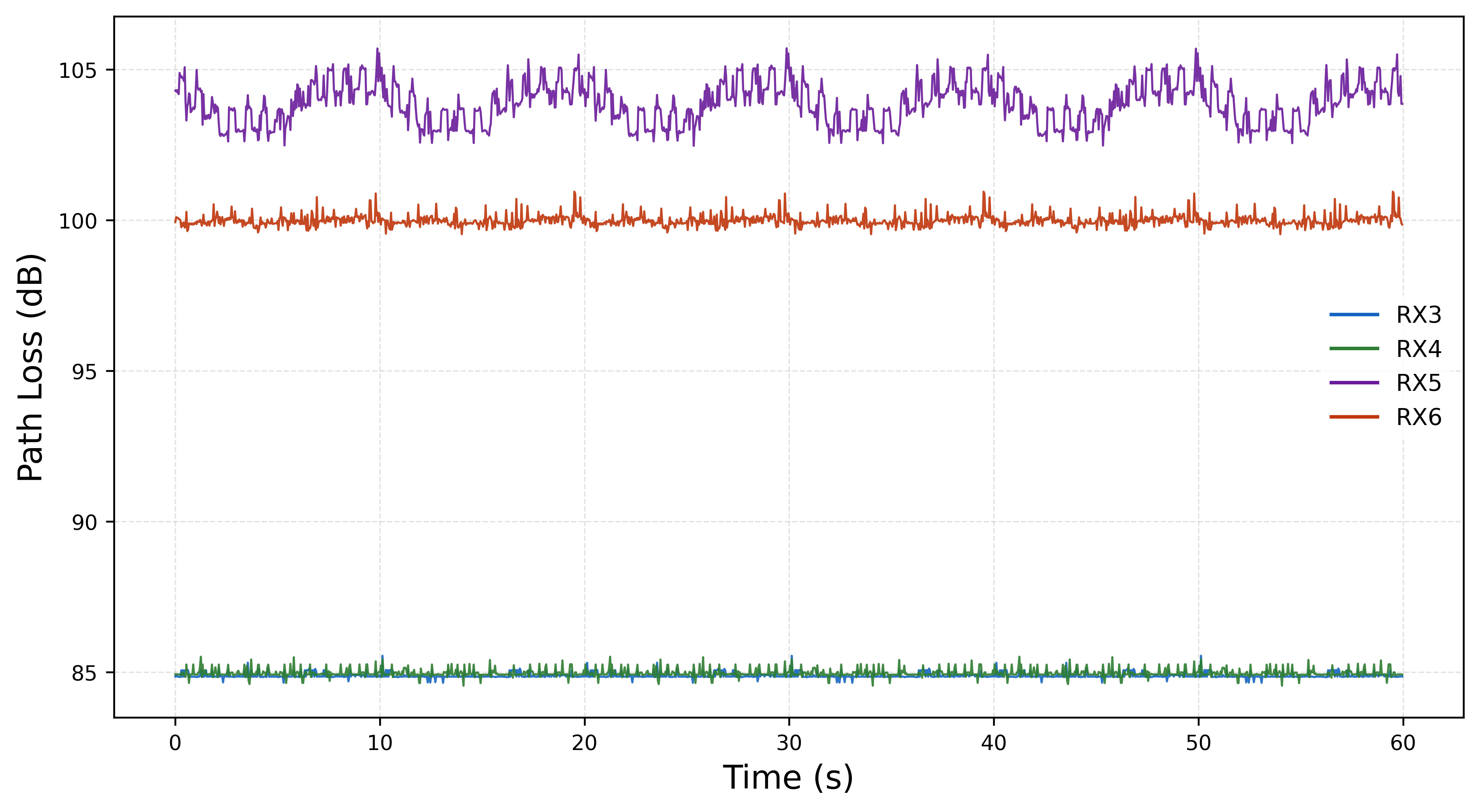}
        \caption{PL variation.}
        \label{PL}
    \end{subfigure}
    \hfill
    \begin{subfigure}[t]{0.49\columnwidth}
        \centering
        \includegraphics[width=\linewidth]{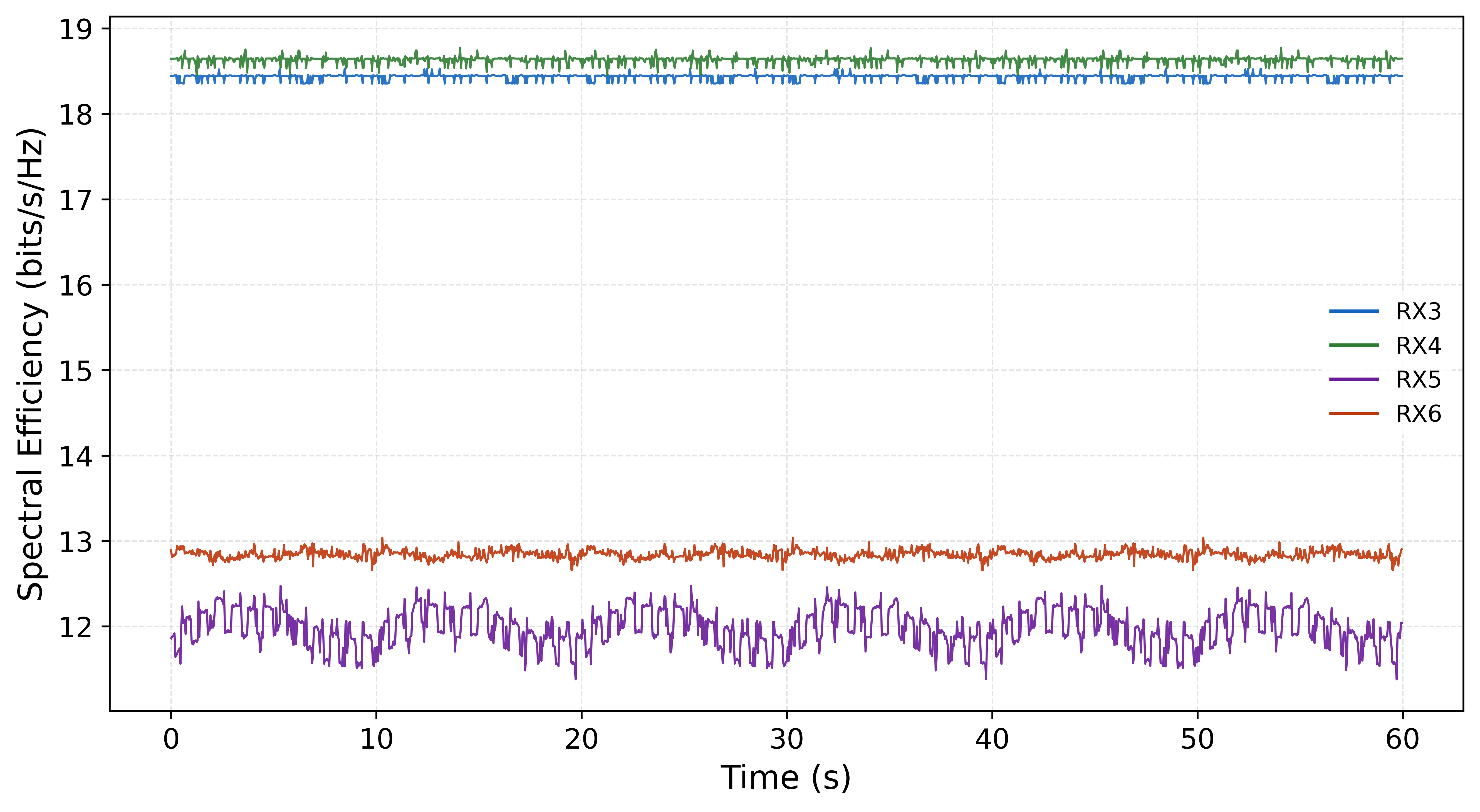}
        \caption{SE variation.}
        \label{SE}
    \end{subfigure}

    \caption{Impact of chest motion on (a) PL and (b) SE at 20 GHz, during measurement.}
    \label{fig:pl_se_20GHz}
\end{figure}

In Fig. \ref{PL}, the variation of PL at 20 GHz across the measurement duration for different communication nodes is visualized. For RX5, it can be observed that PL follows a periodic pattern, which is similar to the periodicity of the ground-truth breathing. The perturbations however, can be attributed to beating heart and appearance and disappearance of the paths. RXs 3, 4 and 6, however do not capture the breathing variations. RXs 3 and 4 are located behind the human and do not have strong contributions from nLOS paths. For RX6, even though it is placed in front of the human, with clear LOS, the paths combine destructively leading to failure in capturing the breathing signal. Similar observations can be made for Fig. \ref{SE}, where SE varies with the chest motion. 

This confirms that positioning RXs strategically in a scene can be utilized for passive sensing through metrics like PL and SE. For RX5 however, it is imperative to mention that it follows an inverse relationship with chest motion, which indicates that during inhalation, the paths reflected off the chest combine coherently, reducing PL, and vice-versa. SE however directly correlates to the breathing waveform.

\section{Conclusion and Future Work}
\label{conclusion}
The goal of this study was to build a RT-based framework for vital signs estimation in the FR3 band in a multipath-rich environment. Hence, such an environment with an animated human model with breathing and heartbeat signals was simulated in SionnaRT at FR3 frequencies. The simulation results show that extraction of vital signs is feasible in FR3 using a signal processing pipeline on channel data. The performance at FR3 frequencies was compared against representative FR1 and FR2 frequencies. It was observed that estimation errors were not only a product of system parameters but also depend on geometry and propagation characteristics. Hence, advanced signal processing and ML-based algorithms are essential. The communication performance was evaluated at communication nodes and it was found that chest motion caused variations in communication metrics, which in turn can be used for passive sensing at favorable positions. Additionally, it can be observed that the FR3 band simulations show promising ISAC performance, providing a good balance between sensing and communication.

This work also highlights the importance and modularity of simulation models for generating realistic channel data, which can be utilized to address the gap of good, reliable and diverse datasets. However, more effort is required in studying the scattering and penetration properties of the human body and various clothing at FR3 frequencies and ISAC performance under varying BR and HR states. The feasibility of multi-band vital signs sensing will be investigated. It is also important to recognize that the current simulation framework can not be a direct replacement of in-situ measurements, and will be validated against real world measurements in future work, to build a true Digital Twin.
\balance
\bibliographystyle{IEEEtran}
\bibliography{references}
\vspace{12pt}





\end{document}